\documentstyle[12pt]{article}
\def\one{1\hskip-.37em}
\def\l{\lambda}
\def\D{{\cal D}}
\def\K{{\cal K}}
\def\H{{\cal H}}
\def\E{{\rm E}\hskip-.55em{\rm I}\,}

\def\ir{{\rm I}\hskip-.2em{\rm R}}
\def\half{\textstyle{\frac{1}{2}}}
\def\quarter{\textstyle{\frac{1}{4}}}

\def\ra{\rightarrow}
\def\tint{{\textstyle{\int}}}
\def\d{\partial}
\def\o{\overline}
\def\b{\begin{eqnarray*}}     
\def\e{\end{eqnarray*}}       
\def\bn{\begin{eqnarray}}     
\def\en{\end{eqnarray}}       
\def\<{\langle}
\def\>{\rangle}

\def\{{\lbrace}
\def\}{\rbrace}
\title{Quantization of Systems with Constraints\footnote{Based on a 
presentation at the International Symposium Symmetries in Science IX, 
Bregenz, Austria, August 6-10, 1996. }}
\author{John R. Klauder\\
Departments of Physics and Mathematics\\
University of Florida\\
Gainesville, Fl  32611}
\date{}
\begin{document}
\maketitle
\begin{abstract}
New measures for the quantization of systems with constraints are discussed 
and applied to several examples, in particular, examples of alternative but 
equivalent formulations of given first-class constraints, as well as a 
comparison of both regular and irregular constraints.
\end{abstract}
\section{Introduction}
The quantization of systems with constraints is of considerable importance in 
a variety of applications. 
Let $\{p_j,q^j\}$, $1\leq j\leq J$, denote a set of dynamical variables, 
$\{\l^a\}$, $1\leq a\leq A\leq 2J$, a set of Lagrange multipliers, and 
$\{\phi_a(p,q)\}$ a set of constraints. Then the dynamics of a constrained 
system may be summarized in the form of an action principle by means of the 
classical action (summation implied)
\bn I=\tint[p_j{\dot q}^j-H(p,q)-\l^a\phi_a(p,q)]\,dt\;.  \en
The resultant equations of motion that arise from the action read
\bn  &&{\dot q}^j=\frac{\d H(p,q)}{\d p_j}+\l^a\frac{\d\phi_a(p,q)}{\d p_j}
\equiv\{q^j,H\}+\l^a\{q^j,\phi_a\}\;,\nonumber\\
&&{\dot p}_j=-\frac{\d H(p,q)}{\d q^j}-\l^a\frac{\d\phi_a(p,q)}{\d q^j}\equiv
\{p_j,H\}+\l^a\{p_j,\phi_a\}\;,\nonumber\\
&&\phi_a(p,q)=0\;,  \en
where $\{\cdot,\cdot\}$ denotes the Poisson bracket. The set of conditions 
$\{\phi_a(p,q)=0\}$ define the {\it constraint hypersurface}. If the 
constraints satisfy 
\bn &&\{\phi_a(p,q),\phi_b(p,q)\}=c_{ab}^{\;\;\;\;c}\,\phi_c(p,q)\;,\\
&&\{\phi_a(p,q),H(p,q)\}=h_a^{\;\;b}\,\phi_b(p,q)\;,  \en
then we are dealing with a system of first-class constraints. If the 
coefficients $c_{ab}^{\;\;\;\;c}$ and $h_a^{\;\;b}$ are constants, then 
it is a system of closed first-class constraints; if they are suitable 
functions of the variables $p$ and $q$, then it is a system of open 
first-class constraints. If the first or both of the conditions in (3) and 
(4) fail, then the system involves second-class constraints. 

For first-class constraints it is sufficient to impose the constraints at 
the initial time inasmuch as the equations of motion will ensure that the 
constraints are fulfilled at all future times. Such an initial imposition 
of the constraints is called an {\it initial value equation}. Furthermore, 
the Lagrange multipliers are not determined by the equations of motion; 
rather they must be specified (a choice of ``gauge'') in order for a 
solution of the dynamical equations to be given. For second-class 
constraints, on the other hand, the Lagrange multipliers are determined 
by the equations of motion in such a way that the constraints are satisfied 
for all time. 

In the remainder of this section we briefly review standard quantization 
procedures for systems with closed first-class constraints, both of the 
operator and path integral variety, pointing out some problems along the way. 
In the following section we develop our coherent state approach for closed 
first-class constraints, which are illustrated by examples in the final 
section. For further details of such general systems, as well as a discussion 
of open first-class and second-class constraints, see Ref.~\cite{kla}.
\subsubsection*{Standard operator quantization}
For a system of closed first-class constraints we assume (with $\hbar=1$) 
that 
\bn &&[\Phi_a(P,Q),\Phi_b(P,Q)]=ic_{ab}^{\;\;\;\;c}\,\Phi_c(P,Q)\;,\\
&&[\Phi_a(P,Q),{\cal H}(P,Q)]=ih_a^{\;\;b}\,\Phi_b(P,Q)\;,  \en
where $\Phi_a$ and $\H$ denote self-adjoint constraint and Hamiltonian 
operators, respectively. Following Dirac \cite{dir}, we adopt the 
quantization prescription given by
\bn i{\dot W}(P,Q)=[W(P,Q),{\cal H}(P,Q)]  \en
where $W$ denotes any function of the kinematical operators $\{Q^j\}$ and 
$\{P_j\}$ which are taken as a self-adjoint, irreducible representation of 
the commutation rules $[Q^j,P_k]=i\delta^j_k\one$, with all other commutators 
vanishing. The equations of motion hold for all time $t$, say $0<t<T$. On the 
other hand, the conditions
\bn  \Phi_a(P,Q)|\psi\>_{\rm phys}=0  \en
that determine the physical Hilbert space are imposed only at time $t=0$ as 
the analog of the initial value equation; the quantum equations of motion 
ensure that the constraint conditions are fulfilled for all time.

The procedure of Dirac has potential difficulties if zero lies in the 
continuous spectrum of the constraint operators for in that case there are 
no normalizable solutions of the constraint condition (8). We face the same 
problem, of course, and our resolution is discussed below; see also 
Ref.~\cite{kla}.
\subsubsection*{Standard path integral quantization}
Faddeev \cite{fad} has given a path integral formulation in the case of 
closed first-class constraint systems as follows. The formal path integral
\bn &&\int\exp\{i\tint_0^T[p_j{\dot q}^j-H(p,q)-\l^a\phi_a(p,q)]\,dt\}\,
\D p\,\D q\,\D\l \nonumber\\
&&\hskip1.5cm=\int\exp\{i\tint_0^T[p_j{\dot q}^j-H(p,q)]\,dt\}\,\delta
\{\phi(p,q)\}\,\D p\,\D q \en
may well encounter divergences in the remaining integrals. Therefore, 
subsidiary conditions in the form $\chi^a(p,q)=0$, $1\leq a\leq A$, are 
imposed picking out (ideally) one gauge equivalent point per gauge orbit, 
and in addition a  factor (in the form of a determinant) is introduced to 
formally preserve canonical covariance. The result is the path integral 
\bn  \int\exp\{i\tint_0^T[p_j{\dot q}^j-H(p,q)]\,dt\}\,\delta\{\chi(p,q)\} 
\det(\{\chi^a,\phi_b\})\delta\{\phi(p,q)\}\,\D p\,\D q\,.  \en
This result may also be expressed as
\bn \int\exp\{i\tint_0^T[p^*_j{\dot q}^{*j}-H^*(p^*,q^*)]\,dt\}\,\D p^*\,
\D q^*\;,  \en
namely, as a path integral
over a reduced phase space in which the $\delta$-functionals have been used 
to eliminate $2A$ integration variables. 

The final expression generally involves curvilinear phase-space coordinates 
for which the definition of the path integral is typically ill defined. 
Additionally, in the form (10), the Faddeev-Popov determinant often suffers 
from ambiguities connected with inadmissible gauge fixing conditions 
\cite{gri}. Thus this widely used prescription is not without its 
difficulties.
\subsubsection*{BRST-BFV formulation}
By extending the phase space to include Grassmann variables, it is possible 
to develop alternative and more powerful methods to discuss systems with 
constraints. These methods are well documented, e.g., \cite{hen}, and will 
not be discussed here.  Instead, our interest focuses on what can be said 
without enlarging the number of variables beyond those that make up the 
original phase space augmented by the necessary Lagrange multipliers.
\section{Coherent State Path Integral}
Canonical coherent states may be defined by the relation
\bn  |p,q\>\equiv e^{-iq^jP_j}\,e^{ip_jQ^j}\,|0\>\;,  \en
where $|0\>$ traditionally denotes a normalized, unit frequency, harmonic 
oscillator ground state. Here $\{Q^j\}$ and $\{P_j\}$, $1\leq j\leq J$, 
denote an irreducible set of self-adjoint operators satisfying the Heisenberg 
commutation relations. The coherent states admit a resolution of unity in 
the form
\bn  \one=\tint\,|p,q\>\<p,q|\,d\mu(p,q)\;,\hskip1.5cm d\mu(p,q)\equiv 
d^J\!p\,d^J\!q/(2\pi)^J\;,  \en
where the integration is over $\ir^{2J}$ and this integration domain and the 
form of the measure are unique. For a general operator ${\cal H}(P,Q)$ we 
introduce the upper symbol
\bn  H(p,q)\equiv\<p,q|{\cal H}(P,Q)|p,q\>=\<p,q|:H(P,Q):|p,q\> \;  \en
which is related to the normal-ordered form as shown. If $\cal H$ denotes 
the quantum Hamiltonian, then we shall adopt $H(p,q)$ as the classical 
Hamiltonian. We note that in the general case $H(p,q)\neq {\cal H}(p,q)$ but 
differ by terms which are $O(\hbar)$. Although these functions may be 
numerically different, in most cases the difference between these two 
functions is qualitatively insignificant. As remarked below, however, there 
are analogous cases where the qualitative difference is quite significant. 
Lastly, we also note that an important one-form is given by
$i\<p,q|d|p,q\>=p_j\,dq^j$. 

Using these quantities, the coherent state path integral for the  
time-dependent Hamiltonian ${\cal H}(P,Q)+\l^a(t)\Phi_a(P,Q)$ is readily 
given by
\bn  &&\<p'',q''|{\sf T}e^{-i\tint_0^T[{\cal H}(P,Q)+\l^a(t)\Phi_a(P,Q)]\,dt}
|p',q'\>\nonumber\\
&&\hskip.6cm=\lim_{\epsilon\ra0}\int\prod_{l=0}^N\<p_{l+1},q_{l+1}|e^{-i
\epsilon({\cal H}+\l^a_l\Phi_a)}\,|p_l,q_l\>\prod_{l=1}^N\,d\mu(p_l,q_l)
\nonumber\\
&&\hskip.6cm=
\int\exp\{i\tint[i\<p,q|(d/dt)|p,q\>-\<p,q|{\cal H}+\l^a\Phi_a|p,q\>]\,dt\}
\,\D\mu(p,q)\nonumber\\
&&\hskip.6cm= {\cal M}\int\exp\{i\tint[p_j{\dot q}^j-H(p,q)-\l^a\phi_a(p,q)]
\,dt\}\,\D p\,\D q \;.  \en
In the second line we have set $p_{N+1},q_{N+1}=p'',q''$ and $p_0,q_0=p',q'$, 
and repeatedly inserted the resolution of unity; in the third and fourth 
lines we have formally interchanged the continuum limit and the integrations, 
and written for the integrand the form it assumes for continuous and 
differential paths ($\cal M$ denotes a formal normalization constant). The 
result evidently depends on the chosen form of the functions $\{\l^a(t)\}$. 
\subsubsection*{Enforcing the quantum constraints}
Let us next introduce the quantum analog of the initial value equation. For 
simplicity we assume that the constraint operators generate a compact group; 
the case of a noncompact group is implicitly discussed below (see 
Ref.~\cite{kla}). In that case 
\bn \E\,\equiv\tint e^{-i\xi^a\Phi_a(P,Q)}\,\delta\xi  \en
defines a {\it projection operator} onto the subspace for which $\Phi_a=0$ 
provided that $\delta\xi$ denotes the normalized, $\tint\delta\xi=1$, group 
invariant measure. Based on (5), (6), and (16) it follows that
\bn && \hskip1.5cm e^{-i\tau^a\Phi_a}\E\,=\E\,\;,  \\
&& e^{-i{\cal H}T}\E\,=\E\,e^{-i{\cal H}T}\E\,=\E\,e^{-i(\E\;{\cal H}\E\;)T}
\E\,\;.  \en
We now project the propagator (15) onto the quantum constraint subspace 
which leads to the following set of relations
\bn &&\hskip-1cm\int\<p'',q''|{\sf T}e^{-i\tint[{\cal H}+\l^a(t)\Phi_a]\,dt}
\,|{\o p}',{\o q}'\>\<{\o p}',{\o q}'|\E\,|p',q'\>\,d\mu({\o p}',{\o q}')
\nonumber\\
&&=\<p'',q''|{\sf T}e^{-i\tint[{\cal H}+\l^a(t)\Phi_a]\,dt}\,\E\,|p',q'\>
\nonumber\\
&&=\lim\,\<p'',q''|[\prod^{\leftarrow}_l(e^{-i\epsilon{\cal H}}e^{-i\epsilon
\l^a_l\Phi_a})]\,\E\,|p',q'\>\nonumber\\
&&=\<p'',q''|e^{-iT{\cal H}}e^{-i\tau^a\Phi_a}\,\E\,|p',q'\>\nonumber\\
&&=\<p'',q''|e^{-iT{\cal H}}\,\E\,|p',q'\>\;,  \en
where $\tau^a$ incorporates the functions $\l^a$ as well as the structure 
parameters $c_{ab}^{\;\;\;\;c}$ and $h_a^{\;\;b}$.
Alternatively, this expression has the formal path integral representation
\bn \int\exp\{i\tint[p_j{\dot q}^j-H(p,q)-\l^a\phi_a(p,q)]\,dt-i\xi^a
\phi_a(p',q')\}\,\D\mu(p,q)\,\delta\xi\;.  \en
On comparing (19) and (20) we observe that  after projection onto the quantum 
constraint subspace the propagator is entirely independent of the choice of 
the Lagrange multiplier functions. In other words, the projected propagator 
is gauge invariant; see Refs.~\cite{kla,hen,sha}.

We may also express the physical (projected) propagator in a more general 
form, namely,
\bn &&\hskip-1cm\int\exp\{i\tint[p_j{\dot q}^j-H(p,q)-\l^a\phi_a(p,q)]\,dt\}
\,\D\mu(p,q)\,\D C(\l)\nonumber\\
&&\hskip.1cm=\<p'',q''|e^{-iT{\cal H}}\,\E\,|p',q'\>  \en 
provided that $\tint\D C(\l)=1$ and that such an average over the functions 
$\{\l^a\}$ introduces (at least) one factor $\E\,$. 
\subsubsection*{Reproducing kernel Hilbert spaces}
The coherent state matrix elements of $\E$ define a fundamental kernel
\bn \K(p'',q'';p',q')\equiv\<p'',q''|\E|p',q'\>\;,  \en
which is a bounded, continuous function
for any projection operator $\E$, especially including the unit operator. It 
follows that $\K(p'',q'';p',q')^*=\K(p',q';p'',q'')$ as well as
\bn \sum_{k,l=1}^K\alpha^*_k\alpha_l\K(p_k,q_k;p_l,q_l)\geq0\en
for all sets $\{\alpha_k\}$, $\{p_k,q_k\}$, and all $K<\infty$. The last 
relation is an automatic consequence of the complex conjugate property and 
the fact that
\bn \K(p'',q'';p',q')=\int \K(p'',q'';p,q)\,\K(p,q;p',q')\,d\mu(p,q)\en
holds in virtue of the coherent state resolution of unity and the properties 
of $\E$. The function $\K$ is called the {\it reproducing kernel} and the 
Hilbert space it engenders is termed a {\it reproducing kernel Hilbert 
space}. A dense set of elements in the Hilbert space is given by functions of 
the form
\bn \psi(p,q)=\sum_{k=1}^K \alpha_k\K(p,q;p_k,q_k)\;,  \en
and the inner product of this function with itself has two equivalent forms 
given by
\bn &&\hskip-1.2cm (\psi,\psi)=\sum_{k,l=1}^K\alpha^*_k\alpha_l\K(p_k,q_k;
p_l,q_l)\\
&&=\tint\psi(p,q)^*\psi(p,q)\,d\mu(p,q)\;.  \en
As usual, the inner product of two distinct functions may be determined by 
polarization. Clearly the entire Hilbert space is characterized by the 
reproducing kernel $\K$. Change the kernel $\K$ and one changes the 
representation of the Hilbert space. Following a suitable limit, it is even 
possible to change the {\it dimension} of the Hilbert space, as we discuss 
in the next section.
\subsubsection*{Reduction of the reproducing kernel}
Suppose the reproducing kernel depends on a number of variables and 
additional parameters. We can generate new reproducing kernels from a given 
one by a variety of means. For example, the expressions
\bn &&\K_1(p'';p')=\K(p'',c;p',c)\;,\\
&&\K_2(p'';p')=\tint w(q'')^*w(q')\K(p'',q'';p',q')\,dq''\,dq'\;,\\
&&\K_3(p'',q'';p',q')=\lim\K(p'',q'';p',q')  \en
each generate a new reproducing kernel provided the resultant function 
remains continuous. Sometimes, however, the inner product in the Hilbert 
space generated by the new reproducing kernel is only given by an analog of 
(26) and not by (27), although frequently some sort of local integral 
representation for the inner product may also exist.

Let us present an example of the reduction of a reproducing kernel. Consider 
the example
\bn &&\hskip-.4cm\<p'',q''|\E|p',q'\>\nonumber\\
&&\hskip.4cm=\pi^{-1/2}\int_{-\delta}^\delta\exp[-\half(k-p'')^2+ik(q''-q')-
\half(k-p')^2]\,dk\;,  \en
which defines a reproducing kernel for any $\delta>0$ that corresponds to an 
infinite dimensional Hilbert space. Let us multiply this expression by 
$\pi^{1/2}/(2\delta)$ and take the limit $\delta\ra0$.
The result is the expression
\bn \K(p'';p')=e^{-\half(p''^2+p'^2)}\;,  \en
which has become a reproducing kernel that characterizes a 
{\it one}-dimensional Hilbert space with every functional representative 
proportional to $\chi(p)\equiv\exp(-p^2/2)$. This one-dimensional Hilbert 
space representation also admits a local integral representation for the 
inner product given by
\bn  (\chi,\chi)=\tint |\chi(p)|^2\,dp/\sqrt{\pi}\;.  \en

This example is an important one inasmuch as it shows how a constraint 
operator with a continuous spectrum is dealt with in the coherent state 
approach.
\section{Applications}
\subsubsection*{Example 1}
The following example is based on Problem 5.1 in Ref.~\cite{hen}. Consider 
the two-degree of freedom system with vanishing Hamiltonian described by the 
classical action
\bn I=\tint(p_1{\dot q}_1+p_2{\dot q}_2-\l_1p_1-\l_2p_2)\,dt\;.  \en
For notational convenience all indices have been placed as subscripts.
The equations of motion become 
\bn {\dot q}_j=\l_j\;,\hskip.8cm{\dot p}_j=0\;,\hskip.8cm p_j=0\;,\hskip1.2cm 
j=1,2\;.\en
Evidently the Poisson bracket $\{p_1,p_2\}=0$.

As a second version of the same dynamics, consider the classical action
\bn  I=\tint(p_1{\dot q}_1+p_2{\dot q}_2-\l_1p_1-\l_2e^{cq_1}p_2)\,dt\;,  
\en
with $c$ a constant, which leads to the equations of motion
\bn {\dot q}_1=\l_1\;,\hskip.3cm{\dot q}_2=\l_2e^{cq_1}\;,\hskip.3cm{\dot p}_1
=-c\l_2e^{cq_1}p_2\;,\hskip.3cm{\dot p}_2=0\;,\hskip.3cm p_1=e^{cq_1}p_2=0\;.
\en
Since $e^{cq_1}p_2=0$ implies that $p_2=0$, it follows that the two 
formulations are equivalent despite the fact that in the second case
$\{p_1,e^{cq_1}p_2\}=-ce^{cq_1}p_2$, which has a completely different 
algebraic structure when $c\neq0$ as compared to $c=0$.

Let us discuss these two examples from the point of view of a coherent state 
quantization. For the first version we consider
\bn {\cal M}\int\exp[i\tint(p_1{\dot q}_1+p_2{\dot q}_2-\l_1p_1-\l_2p_2)\,dt]
\,\D p\,\D q\,\D C(\l)\;,  \en
which is defined in a fashion to yield
\bn \<p'',q''|\E|p',q'\>\;,\en
where
\bn  \E=\E(-\delta<P_1<\delta)\E(-\delta<P_2<\delta)\;.  \en
In particular this leads to the fact that
\bn &&\<p'',q''|\E|p',q'\>\nonumber\\
&&\hskip-1cm=\pi^{-1}\prod_{l=1}^2\int_{-\delta}^\delta \exp[-\half
(k_l-p''_l)^2+ik_l(q''_l-q'_l)-\half(k_l-p'_l)^2]\,dk_l\;. \en
Let us reduce this reproducing kernel by multiplying this expression by 
$\pi/(2\delta)^2$ and passing to the limit $\delta\ra0$. The result is the 
reduced reproducing kernel given by
\bn \exp[-\half(p''^2_1+p''^2_2)]\,\exp[-\half(p'^2_1+p'^2_2)]\;, \en
which clearly characterizes a particular representation of a one-dimensional 
Hilbert space in which every vector is proportional to $\exp[-\half
(p_1^2+p_2^2)]$. This example is, of course, related to the reduction example 
given earlier. Moreover, we can introduce a local integral representation 
over the remaining $p$ variables for the inner product if we so desire.

Let us now turn attention to the second formulation of the problem by 
focusing (for a different $C(\l)$) on
\bn {\cal M}\int\exp[i\tint(p_1{\dot q}_1+p_2{\dot q}_2-\l_1p_1-\l_2e^{cq_1}
p_2)\,dt]\,\D p\,\D q\,\D C(\l)\;.  \en
This expression again leads (for a different $\E$) to 
\bn \<p'',q''|\E|p',q'\>\;,\en
where in the present case the fully {\it reduced} form of this expression is 
proportional to
\bn &&\hskip-1cm\int\exp[-\half(k_2-p''_2)^2+ik_2(q''_2-q'_2)-\half
(k_2-p'_2)^2]\nonumber\\
&&\times\exp[-\half(k_1-p''_1)^2+ik_1q''_1-\half i\l_1k_1]\nonumber\\
&&\times\exp[-ixk_1-i\l_2e^{cx}k_2+ix\kappa_1]\nonumber\\
&&\times\exp[-\half i\l_1\kappa_1-i\kappa_1q'_1-\half(\kappa_1-p'_1)^2]
\nonumber\\
&&\hskip1cm \times 
dk_2\,dk_1\,dx\,d\kappa_1\,d\l_1\,d\l_2\;.  \en
When normalized appropriately, this expression is evaluated as
\bn \exp[-\half(p''^2_1+p''^2_2+icp''_1)]\,\exp[-\half(p'^2_1+p'^2_2-icp'_1)]
\;,
\en
which once again represents a one-dimensional Hilbert space.

Thus we have obtained a $c$-dependent family of distinct but equivalent 
quantum representations for the same Hilbert space, reflecting the 
$c$-dependent family of equivalent classical solutions. Observe that in the 
quantum theory, just as in the classical theory, all observable effects are 
independent of $c$.
\subsubsection*{Example 2}
In discussing constraints one often pays considerable attention to the 
regularity of the expressions involved; see \cite{hen}, Sec.~1.1.2. Consider, 
once again, the simple example of a single constraint $p=0$ as illustrated by 
the classical action
\bn  I=\tint(p{\dot q}-\l p)\,dt\;.  \en
The equations of motion read ${\dot q}=\l$, ${\dot p}=0$, and $p=0$. On the 
other hand, one may ask about imposing the constraint $p^3=0$ or possibly 
$p^{1/3}=0$, etc., instead of $p=0$. Let us incorporate several such odd 
(function) examples by studying the classical action 
\bn  \tint (p{\dot q}-\l p|p|^\gamma)\,dt\;,\hskip1cm\gamma>-1\;.  \en
Here the equations of motion include ${\dot q}=\l(\gamma+1)\,|p|^\gamma$ 
which, along with the constraint $p|p|^\gamma=0$, may cause some difficulty 
in seeking a classical solution of the equations of motion, e.g., if 
$\gamma<0$. When $\gamma\neq0$, such constraints are said to be 
{\it irregular}. It is clear from (9) that irregular constraints lead to 
considerable difficulty in conventional phase-space path integral 
approaches.

Let us examine the question of irregular constraints from the point of view 
of a coherent state, phase-space path integral quantization. For any 
$\gamma>-1$, we first observe that the operator $P|P|^\gamma$ is well defined 
by means of its spectral decomposition. Moreover, it follows that
\bn && \hskip-1cm\int e^{-i\xi P|P|^\gamma}{\sin(\delta^{\gamma+1}\xi)\over
\pi\xi}\,d\xi\nonumber\\
&&=\E(-\delta^{\gamma+1}<P|P|^\gamma<\delta^{\gamma+1})\nonumber\\
&&=\E(-\delta<P<\delta)\;.  \en
Thus, from the operator point of view, it is possible to consider the 
constraint operator $P|P|^\gamma$ just as easily as $P$ itself. In 
particular, it follows that
\bn \<p'',q''|\E|p',q'\>
={\cal M}\int\exp[i\tint(p{\dot q}-\l p|p|^\gamma)\,dt]\,\D p\,\D q\,\D 
C_\gamma(\l)\;, \en
where we have appended $\gamma$ to the measure for the Lagrange multiplier 
$\l$ to emphasize the dependence of that measure on $\gamma$. The reduction 
of the reproducing kernel proceeds exactly like the cases discussed earlier, 
and we determine for all $\gamma>-1$ that
\bn \lim_{\delta\ra0}{\sqrt{\pi}\over(2\delta)}\<p'',q''|\E|p',q'\>=e^{-\half
(p''^2+p'^2)}\;,  \en
representative of a one-dimensional Hilbert space. Just like the classical 
theory, note that the ultimate form of the quantum theory is independent of 
$\gamma$. 

It is natural to ask how one is to understand this acceptable behavior for 
the quantum theory for irregular constraints while there are difficulties 
that seem to be present in the classical theory. In the first section we 
discussed the definition of the classical generator as derived from the 
quantum generator. Just like the classical and quantum Hamiltonians, the 
connection between the classical and quantum constraints is given by 
\bn \phi(p,q)\equiv\<p,q|\Phi(P,Q)|p,q\>=\<0|\Phi(P+p,Q+q)|0\>\;.   \en
With this rule we typically find that $\phi(p,q)\neq\Phi(p,q)$ due to the 
fact that $\hbar\neq0$, but the difference between these expressions is 
generally qualitatively unimportant. In certain circumstances, however, that 
difference is qualitatively significant even though it is quantitatively very 
small. Since that difference is $O(\hbar)$ let us explicitly exhibit the 
appropriate $\hbar$-dependence hereafter.
First consider the case of $\gamma=2$. In that case
\bn \<p,q|P^3|p,q\>=\<0|(P+p)^3|0\>=p^3+3\<P^2\>p\;,  \en
where we have introduced the shorthand $\<(\cdot)\>\equiv\<0|(\cdot)|0\>$.
Since $\<P^2\>=\hbar/2$ it follows that for the quantum constraint $P^3$, the 
corresponding classical constraint function is given by $p^3+(3\hbar/2)p$. 
For $|p|\gg\sqrt{\hbar}$, this constraint is adequately given by $p^3$. 
However, when $|p|\ll\sqrt{\hbar}$---{\it as must eventually be the case in 
order to actually satisfy the classical constraint}---then the functional 
form of the constraint is effectively $(3\hbar/2)p$. In short, if the quantum 
constraint operator is $P^3$, then the classical constraint function is in 
fact {\it regular} when the constraint vanishes. 

A similar discussion holds for a general value of $\gamma$. The classical 
constraint is given by
\bn &&\hskip-1.1cm\phi_\gamma(p)
=({\pi\hbar})^{-1/2}\int(k+p)|k+p|^\gamma e^{-k^2/\hbar}\,dk\nonumber\\
&&=({\pi\hbar})^{-1/2}\int k|k|^\gamma e^{-(k-p)^2/\hbar}\,dk\;. \en
For $|p|\gg\sqrt{\hbar}$ the first line of this expression effectively yields 
$\phi_\gamma(p)\simeq p|p|^\gamma$. On the other hand, for $p\approx 0$, and 
more especially for $|p|\ll\sqrt{\hbar}$, the second line of this 
expression shows that this constraint function vanishes {\it linearly}, speci
fically as $\phi_\gamma(p)\simeq kp$, where 
\bn k\equiv 2({\hbar^\gamma/\pi})^{1/2}\int y^2|y|^\gamma e^{-y^2}\,dy=2(
{\hbar^\gamma/\pi})^{1/2}\Gamma((\gamma+3)/2)\equiv \hbar^{\gamma/2}k_o\;. 
\en
A rough, but qualitatively correct expression for this behavior is given by
\bn  \phi_\gamma(p)\simeq k_op(\hbar+p^2k_o^{-2/\gamma})^{\gamma/2}\;. \en

Thus, from the present point of view, irregular constraints do not arise from 
consistent quantum constraints; instead, irregular constraints arise as 
limiting expressions of certain consistent, regular classical constraints as 
$\hbar\ra0$.

There is one category of irregular constraints that is not covered by the 
foregoing discussion, namely the even constraints, e.g., a classical 
constraint given by $p^2=0$ rather than $p=0$. This case differs from those 
treated above because the classical constraint is strictly nonnegative. 
However, from the operator point of view, the case of $P^2$ is not 
qualitatively different from the other cases because we still have
\bn && \hskip-1cm\int e^{-i\xi P^2}{\sin(\delta^2\xi)\over\pi\xi}\,d\xi
\nonumber\\
&&=\E(P^2<\delta^2)\nonumber\\
&&=\E(-\delta<P<\delta)\;.  \en
From the classical point of view, however, it follows that
\bn  \phi(p)=p^2+\<P^2\>=p^2+\half\hbar\;,  \en
which evidently never vanishes so long as $\hbar>0$. This result is not 
surprising; as the expectation value of a nonnegative operator with 
continuous spectrum, it cannot vanish. If we adopt our previous 
interpretation that there are no irregular odd constraints thanks to a 
nonvanishing $\hbar$, then we are forced to admit that for even constraints 
there is no suitable classical analog. Of course, such examples are in no 
way restricted to terms involving $\hbar$. One need only consider the 
classical constraint $p^2+q^2+1=0$ which evidently has no solution for real 
phase space variables. 

However, there is another way to look at even classical constraints 
\cite{arn}. For the sake of illustration, let us initially focus on the 
simple case $p^2=0$ as an even representative of the odd classical constraint 
$p=0$. 
From a simple physical point of view, a system will follow a constraint at
least approximately provided it costs a great deal of energy to violate it. 
onsider the simple system described by the action functional
  \bn  I=\tint(p{\dot q}-\half Ap^2)\,dt\;,  \en
where $A$ is a large positive constant. The equations of motion for this case 
read ${\dot q}=Ap$, and ${\dot p}=0$, and the solution to these equations of 
motion is $p(t)=p'$ and $q(t)=Ap't+q'$. We insist that the solution should be 
independent of the large parameter $A$, and so it follows that 
$p'=O(A^{-1})$. As a consequence, the ``energy'' $Ap'^2/2=O(A^{-1})$ as well. 
In the limit that $A\ra\infty$, it follows that $p'=0$, the ``energy'' 
vanishes, while $q(t)=ct+q'$ for some $c$. This solution is seen to be an 
example of one from the usual formulation using a Lagrange multiplier. 
Indeed, we can also make $q(t)$ into a rather general function of time by 
allowing for $A$ to be time dependent as well as large; however, for 
convenience, we ignore this rather evident generalization. An alternative 
example of the same kind is given by
 \bn I=\tint[p{\dot q}-\half A(p^2+q^2-1)^2]\,dt\;.  \en
The analysis in this case proceeds as before with the final result that 
 \bn  q(t)=q'\cos(ct)+p'\sin(ct)\;.\hskip1cm p(t)=-q'\sin(ct)+p'\cos(ct)\;,
\en
subject to the requirement that $p'^2+q'^2=1$. Note well, in these cases, 
that we deal with even constraints and a large positive parameter $A$ in 
order that no cancellation among large terms of opposite sign is possible.

Following this brief introduction to an alternative classical formulation of 
a theory with a constraint, let us turn our attention to the quantum 
mechanics of such a system. Let $W$ denote the basic constraint operator we 
wish to maintain; for instance, $W=P^2/2$ for the first case and 
$W=(:P^2+Q^2:-1)^2/2$ for the second case. Maintaining $\hbar=1$ temporarily, 
interest centers on the expression
 \bn \lim K_A e^{-iTAW}  \en
as $A\ra\infty$, where $K_A$ is a suitable $A$-dependent parameter. If $W=0$ 
is a point in the discrete spectrum, as in the second example, then $K_A=1$ 
and the result is a projection operator $\E$ onto the subspace where $W=0$. 
The case of a discrete spectrum is not especially difficult and we choose to 
focus our attention on the first example for which $W=0$ is a point in the 
continuous spectrum. In that case it follows that
 \bn &&\lim_{A\ra\infty}\sqrt{iAT/2\pi}\;\<p'',q''|e^{-iTAP^2/2}|p',q'\>
\nonumber\\
&&\hskip.5cm=\lim_{A\ra\infty}\sqrt{iAT/2\pi}\;\int \exp[-\half(k-p'')^2
\nonumber\\
&&\hskip2.5cm+ik(q''-q')-i\half TAk^2-\half(k-p')^2]\,dk\nonumber\\
&&\hskip.5cm=\exp[-\half(p''^2+p'^2)]\;,  \en
which is exactly the same result that we would find had we started with the 
usual formulation with a Lagrange multiplier. 

Let us now present a coherent state path integral for this version of doing 
constraint systems. In particular we observe that
\bn  \<p'',q''|e^{-iTAP^2/2}|p',q'\>={\cal M}\int\exp\{i\tint[p{\dot q}-\half 
A(p^2+\<P^2\>)]\,dt\}\,\D p\,\D q\;.  \en
In this formal path integral we have included the term
  \bn  \<p,q|P^2|p,q\>=p^2+\<P^2\>=p^2+\half\hbar\;.  \en
Thus, from the point of view of deriving the classical action from the 
quantum theory, we are not led to the original starting action (59) but 
instead to
  \bn  I=\tint[p{\dot q}-\half A(p^2+\half\hbar)]\,dt \;. \en

When a term such as (65) arose as the coefficient of a Lagrange multiplier, 
we concluded that it was unacceptable because it was impossible to satisfy 
$p^2+\hbar/2=0$ as a constraint for real phase space variables. On the other 
hand, we now see such a term arise as a coefficient of the large parameter 
$A$. Note that the extra $\hbar$ term does {\it not} influence the equations 
of motion. In particular, the classical equations of motion are ${\dot q}=Ap$ 
and ${\dot p}=0$ exactly as before. Thus the criterion that the classical 
solutions should be independent  of the large parameter $A$ leads once again 
to the solutions $p(t)=0$ and $q(t)=ct+q'$. It is true that the ``energy'' 
given by $A(p'^2+\hbar/2)/2=A\hbar/4$ diverges as $A\ra\infty$, but that 
expression is not the real energy for any physical system. Thus we conclude 
that the appearance of the extra term in the classical action does not 
interfere with an appropriate solution of the classical equations of motion 
unlike the formulation with a Lagrange multiplier. 

Let us briefly raise the issue in the case of another even constraint, say 
$p^4=0$ based on the classical action
  \bn I=\tint[p{\dot q}-\quarter Ap^4]\,dt\;.  \en
The equations of motion, ${\dot q}=Ap^3$ and ${\dot p}=0$, lead to an 
$A$-independent form of the classical solution given by $q(t)=ct+q'$, for 
some $c$, and $p(t)=0$. Quantum mechanically, we are led to conclude that
 \bn \lim_{A\ra\infty}kA^{1/4}\<p'',q''|e^{-iTAP^4/4}|p',q'\>=\exp[-\half
(p''^2+p'^2)]\;,  \en
for a suitable constant $k$. The coherent state path integral formulation 
leads to
 \bn  {\cal M}\int \exp\{i\tint[p{\dot q}-\quarter A(p^4+6\<P^2\>p^2+
\<P^4\>)]\,dt\}\,\D p\,\D q\;,  \en
which implies that the proper factor in the classical theory
is
 \bn \quarter A[p^4+3\hbar p^2+(3/4)\hbar^2]\;.  \en
If we take this form seriously, we are led to conclude that as $A\ra\infty$ 
and $|p|\ll\sqrt{\hbar}$, only the quadratic term becomes important in the 
equations of motion as compared with the quartic term. In short, the $P^4$ 
term is just as ``regular'' as the $P^2$ term. Thus, once again, we conclude 
that from the coherent state point of view there are no irregular constraints 
so long as $\hbar>0$; it is only in the limit where $\hbar\ra0$ that certain 
regular constraints turn into irregular constraints.

\section*{Acknowledgements}
Thanks are expressed to Bruno Gruber for his efforts which resulted in a very 
pleasant conference. Communications with J. Govaerts, M. Henneaux, and S. 
Shabanov, and discussions with B. Whiting are gratefully acknowledged.

\end{document}